\documentclass[preprint,showpacs,showkeys,preprintnumbers]{revtex4}

\usepackage{graphicx}% Include figure files

\begin{document}

\preprint{}

\title{A discrete nonlinear mass transfer equation with applications in 
\\solid-state sintering of ceramic materials} 
% Force line breaks with \\
\author{Dionissios T. Hristopulos}
\email{dionisi@mred.tuc.gr } 
 \homepage{http://www.mred.tuc.gr/dionisi.htm}
 \author{ Leonidas Leonidakis}
 \thanks{Mr. Leonidakis was killed in a car accident around Chania on August 6, 2005}
 \author{Athena Tsetsekou}
\email{athtse@mred.tuc.gr }
\affiliation{Department of Mineral Resources Engineering\\
Technical University of Crete\\Chania 73100, Greece}%
\thanks{}
\begin{abstract}
The evolution of grain structures in materials is a complex and multiscale process that determines the material's final properties \cite{gold05}.  
Understanding the dynamics of grain growth is a key factor for controlling this process.  
We propose a phenomenological approach, based on a nonlinear, discrete mass transfer equation for the evolution of an arbitrary initial grain size distribution.  
Transition rates for mass transfer across grains 
are assumed to follow the Arrhenius law, but
the activation energy depends on the degree of amorphization of
each grain. 
We argue that the magnitude of the activation energy controls the 
final (sintered) grain size distribution, and we verify this prediction
by numerical simulation of mass transfer in a one-dimensional 
grain aggregate.
\end{abstract}
\pacs{81.07.Bc, 81.10.Aj, 81.20.Ev } 
\keywords{sintering kinetics, grain size distribution, activation}
\maketitle

\section{Introduction}
Ceramic materials have applications in many fields of science and engineering
\cite{chiang}.  
In particular, the efficient production of dense ceramics with 
nanosized grains using ultrafine powders is a subject of increased
attention.  
The most common method for achieving densification in ceramics
 is solid-state sintering, 
a thermally activated process that involves the diffusion and 
redistribution of mass by means 
of various physical processes, leading 
to a compactified and consolidated 
final structure.  
In the case of nanosized powder compacts, a major concern
is the considerable grain growth during the densification 
process. Special consolidation techniques such as hot isostatic pressing, spark plasma and microwave sintering 
in many cases successfully limit this undesirable phenomenon. 
The addition of dopants able to limit grain boundary migration is another effective option. 

A very effective way to limit grain growth is to promote densification at a lower sintering temperature \cite{palm}. 
Mechanical activation of powders by high-energy ball milling can contribute greatly toward this direction. High-energy mechanical treatment of solids usually leads to the accumulation of excess energy in the structure resulting in plastic deformation and fracturing of crystals \cite{bokh,juha}. The process usually involves initial reduction of grain sizes, followed by plastic deformation of the crystalline structure, whereas excessive milling time may finally lead to grain aggregation  \cite{bal,bokh,bold2,kostic}. 
Grain size reduction leads to an increase of surface energy 
\cite{stoj}. 
Under deformation the lattice structure is transformed, perturbed or completely destroyed. This effect is evidenced by the 
broadening, decrease of intensity, and finally total disappearance of XRD reflections \cite{juha}.
This means that the deformation and increase of structural disorder can bring the solid into a nanocrystalline or even completely amorphous state [6, 11-13]. Finally, the possible aggregation of grains imposes significant pressure on the grains increasing their contact surface areas \cite{bal,bold2}.
This lowers the activation energy for the diffusion processes 
involved in sintering, thus accelerating the entire process
 and reduces the sintering 
temperatures \cite{stoj,heegn}.

The alteration of the state of solid materials due to 
mechanical activation can lead to significant changes in physical and chemical properties. 
Order-disorder transitions and phase transformations can then take place at much lower temperatures. Additionally, nuclei of new phases can often appear at the interfaces possibly followed by subsequent growth of crystallites 
\cite{bal,kostic,xue}. 
Thus, crystallization from the amorphous state 
and chemical reactions can be easily accelerated with subsequent treatment, resulting in materials with improved or better-controlled properties 
\cite{stoj,xue}.

Therefore, a central issue for mechanically 
activated powders is the relations between the 
properties of the initial powder and the structure of the resulting ceramic material. 
Improved performance is in many cases achieved by an optimal balance between particle size and mechanical activation effects \cite{bold1}. 
Theoretical and computational investigations can help to understand the impact of mechanical activation on the sintering behaviour of nanosized powders ,
as well as on sintering temperature dependence. 

Even though theories of the sintering process has been intensely 
researched during the last fifty years, 
 there is a lack of quantitative, predictive models for the 
microstructural evolution during sintering \cite{chiang}.
In the final stage of sintering the densification 
process is accompanied by grain growth leading to
the final pore structure (grain size distribution, geometry and topology 
of pores).  The physical mechanisms responsible for these effects involve
various transport processes acting in the bulk and on the surface of grains, 
 at the boundary between grains, as well as the evaporation-condensation
mechanism.   

Given the complexity of the physical processes and the initial micro-structure, 
it is not surprising that a variety of mathematical
models have been proposed that focus on different stages of the sintering process, based on 
various simplifying assumptions regarding the number of grains, the geometry and 
topology of the pore structure, etc.
The existing models include continuum diffusion or transport 
equations \cite{olev98,kuch00,maxim04} that are solved either analytically 
(in simplified cases) or numerically, and
kinetic  Monte Carlo (Potts model) simulation approaches \cite{brag}. 
More specifically in the case of grain growth, computational approaches 
involve vertex, Monte Carlo Potts, phase field, and cellular automata models
\cite{miod02}. 

\section{A Nonlinear Equation for Grain Growth}
\label{sde-gg}
We introduce a model of grain growth 
applying to the final stage of sintering. The model 
describes mass transfer between grains in terms of a balance equation.
Such a model should
account for differences in the initial distribution
of grain sizes, which contribute to differential growth 
of the grains.  A stochastic aspect is introduced in 
the problem, due to the fact that the initial grain configuration is not 
known. 
Instead of attempting to model explicitly the various physical mechanisms that contribute to 
mass transfer, we opt for an effective approach that 
does not distinguish between the different mechanisms.
The amorphization degree represents the fraction of the grain
that is in the amorphous state. Grains with higher 
degree of amorpization are more active, and tend to 
transfer mass more effectively. 
Smaller grains 
(e.g., such as those created by mechanical activation) 
are expected to have a higher degree of amorphization.

Since the lattice formed by the centres of the grains is 
irregular, the approach to a continuum limit is not 
well-defined. Thus, a discrete mass transfer equation is proposed.
The considerations in the paragraph above imply that 
transition rates should depend on  
grain size.  
The locations of the grain centers are
denoted by ${\bf x}$, and those of the nearest neighbors 
of point ${\bf x}$ by ${\bf x'}$. The grain masses at 
time $t$ are denoted by $m({\bf x},t)$ and $r({\bf x},t)$ is the grain radius.
The grain mass density (assumed constant) is denoted by $\rho$. 
We introduce a local `order factor' $\phi({\bf x},t)$, 
which is inversely proportional to the `activation' of each grain.  
We assume that the order factor increases with the grain mass, 
since the fraction of amorphous areas 
is smaller for larger grains.  The order factor influences the 
local activation energy.
To our knowledge, there are no experimental measurements of 
activation energies as a function of grain amorpization. 
Hence, it is necessary to 
surmise the functional dependence of the order factor 
based on intuition and 
physical constraints. 
Based on these considerations, we propose the 
following system of equations: 

\begin{equation}
\label{mass-trans}
\partial_{t}\, m({\bf x},t) = \sum_{{\bf x'}} W({\bf x'} \rightarrow {\bf x},t) \,
 m({\bf x'},t) - \sum_{{\bf x'}} W({\bf x} \rightarrow {\bf x'},t) \, m({\bf x},t),
\end{equation}

\begin{equation}
\label{order}
\phi({\bf x},t)= \frac{ \alpha \, m({\bf x},t) }{m_{0}},
\end{equation}

\begin{equation}
\label{rate}
 W({\bf x} \rightarrow {\bf x'},t) = \lambda \, \exp 
   \left\{ - \frac{Q \, \phi ({\bf x},t) } {R \, T} \right\}.
\end{equation}

\noindent
In Eq.~(\ref{mass-trans}) $W({\bf x'} \rightarrow {\bf x},t)$ 
represents the transition rate at time $t$ for mass transfer 
into the grain located at ${\bf x}$ from the neighbouring grains. 
Similarly, $W({\bf x} \rightarrow {\bf x'},t)$ represents the 
transition rate at time $t$ for 
mass transfer from the grain at ${\bf x}$ into the neighbouring grains.
The mass and the radius are related via 
$m({\bf x},t)= g_{n} \, \rho \, r^{n}({\bf x},t),$
where $g_{n}$ is a geometric factor that  
on the dimension $n$. 
The order factor defined by Eq.~(\ref{order}), where
$m_{0}$ is a reference mass and $\alpha$ a dimensionless
constant, satisfies the 
conditions $\phi \rightarrow 0$ for $m \rightarrow 0$ 
and $\phi \rightarrow \infty$ 
for $m \rightarrow \infty$. 
The first condition implies that the activation energy tends to vanish 
 for small grains, which are expected to have a significant 
amorphous fraction, while the second condition implies a very 
large activation energy for big grains that are predominantly ordered. 
In Eq.~(\ref{rate}), $\lambda$ is a constant that 
determines the scale of the transition rates, and
$R=8.31 \, {\rm J}\, {\rm K}^{-1} \, {\rm mol}^{-1}$ is
 the gas constant.  
Based on Eqs.~(\ref{rate}) and (\ref{order}), the following 
equation is obtained for the transition rates:

\begin{equation}
\label{rate2}
 W({\bf x} \rightarrow {\bf x'},t) = \lambda \, \exp 
   \left\{ - \frac{\tilde{Q}} {R \, T} \left[ \frac{m({\bf x},t)}
   {m_{0}} \right]  \right\}.
\end{equation}

Although the mass transfer equation~(\ref{mass-trans}) 
 looks similar to the master equation
used in the analysis of diffusion processes, 
it differs from the latter in the following: First, 
the conserved quantity in the master equation is a probability function,
while in Eq.~(\ref{mass-trans}) it is the total
mass of the grains. Secondly, the classical master equation 
is linear in the probability, while Eq.~(\ref{mass-trans}) is nonlinear 
in the mass due to  
the variations in grain activity. 
Finally, in classical diffusion the 
transitions tend to generate a uniform steady state, 
while this is not necessarily the case  
for Eq.~(\ref{mass-trans}), as shown below.  

For a time-independent steady  state to exist, the time 
derivative of $m({\bf x},t)$ must vanish asymptotically. 
This is accomplished
if the `detailed balance' condition is satisfied, i.e.,

\begin{equation}
\label{det-bal}
\lim_{t \rightarrow \infty} \frac{m({\bf x},t)}{m({\bf x'},t)} =
\lim_{t \rightarrow \infty} \frac{W({\bf x'} \rightarrow {\bf x},t)} 
{W({\bf x} \rightarrow {\bf x'},t)}=\lim_{t \rightarrow \infty}
\exp \left\{-  \frac{ \alpha \, Q }{R \, T \, m_{0}} \, 
\left[  m({\bf x'},t) - 
 m({\bf x},t)    \right] \right\}.
\end{equation}

\noindent
If a steady state exists in which some grains have
$m({\bf x},\infty)=0$, while for at least one of their neighbours  
$m({\bf x'},\infty) \neq 0$, the left hand-side of Eq.~(\ref{det-bal}) tends
to zero. Then  
the right hand-side becomes 
$\exp \left[- \frac{ \alpha \, Q \, m({\bf x'},t)}{R \, T \, m_{0}} \right]$. 
Since $m({\bf x},t) \propto O(m_{0})$, 
a steady state with vanishing grain masses can exist only if
$ \alpha \, Q \equiv \tilde{Q} >>R \, T$.
This condition is necessary but not sufficient for an asymmetric 
steady state.  
Using as an average estimate for the
activation energy the value of $500$ kJ/mol \cite{heegn}
and a typical sintering temperature (e.g., 1500$^{\circ}$ C),
$Q/R\,T \approx  34$, giving an order of magnitude estimate. 
The value of $\alpha$ determines the variation of the activation
energy due to mechanical activation effects. 
While a large value of $Q/R\,T$ is necessary for 
achieving a non-uniform steady state, very large values 
slow down considerably the sintering process. 
We expect that 
$\alpha \leq 1$, to allow for reduction of the activation energy 
by mechanical activation.  
At the opposite limit, if 
$ \alpha \, Q << R \, T$ the ratio of the transition rates 
in Eq.~(\ref{det-bal}) will be 
close to one, and a uniform 
steady-state is expected.

\section{Dimensional Analysis}
\label{dim-anal}

We use the Length (L) - Mass (M) - Time (T) 
system of units for dimensional analysis. 
The governing parameters \cite{bar}  of the grain mass equation are 
$m_{0}, \rho, \lambda, \tilde{Q}, {\bf x}, t$, and $R\, T$. To these one should 
add the parameters that determine the 
initial grain size distribution. If this distribution is approximately 
Gaussian, the parameters include the mean value $\overline{r}_{0}$ 
and standard deviation
$\sigma_{r,0}$.  Then, the governing parameters and their
dimensions are as follows: 
$[m_{0}]=M, \, [\lambda]=T^{-1}, \, [\tilde{Q}]=[R\, T]=M\,L^{2}\,T^{-2}, \,
  [x]=[\overline{r}_{0}]=[\sigma_{r,0}]=L, \, [t]=T, \,
  [\rho]= M\,L^{-3}. $
There are three variables with independent dimensions, i.e., 
$m_{0}, \tilde{Q}, \lambda$. According to the 
$\Pi$-theorem of dimensional analysis \cite{bar}, the equation~(\ref{mass-trans}) 
can be expressed in terms of six dimensionless combinations of 
the governing parameters, e.g., 

\begin{equation}
\label{scaled}
\frac{ m({\bf x},t) } {m_{0} } = \tilde{m} \left( \lambda \, t, \frac{\bf x}{\overline{r}_{0}}; \frac{\tilde{Q}}{R\, T}, \,  \frac{\sigma_{r,0}}{\overline{r}_{0}},
\, \frac{\tilde{Q}}{m_{0}\,\overline{r}_{0}^{2} \, \lambda^{2}}, \,  
  \frac{m_{0}} {\rho \, \overline{r}_{0}^{3}},  \right).
\end{equation}

\noindent
Using the notation  
$\tilde{t}= \lambda \, t$, $\tilde{\bf x} =\frac{\bf x}{r_{0} }$, 
$u=\frac{\tilde{Q}}{R\, T}$, $\mu_{r,0}=\frac{\sigma_{r,0}}{\overline{r}_{0}}$, 
$\tilde{m}_{0} = \frac{m_{0}} {\rho \, \overline{r}_{0}^{3}}$,
$z=\frac{\tilde{Q}}{m_{0}\,\overline{r}_{0}^{2} \, \lambda^{2}}$ 
 for the dimensionless variable combinations, 
 the mass-transfer equation is expressed as follows: 

\begin{equation}
\label{mass-trans2}
\partial_{\tilde{t}}\, \tilde{m}(\tilde{\bf x},\tilde{t}) 
 =    
\sum_{\tilde{\bf x'}} 
e^{  - u  \, \tilde{m}(\tilde{\bf x'},\tilde{t})   }
 \, \tilde{m}(\tilde{\bf x'},\tilde{t}) 
  -   \sum_{\tilde{\bf x'}}  \, 
e^{  - u  \, \tilde{m}(\tilde{\bf x},\tilde{t}) }
 \, \tilde{m}(\tilde{\bf x},\tilde{t}) .
\end{equation}
Note that the dependence on the scaled variables $\mu_{r,0}$, 
$\tilde{m}_{0}$ and $z$
is not explicit in Eq.~(\ref{mass-trans2}), since these variables 
involve the initial conditions.  

\section{Simulations Results and Discussion}
\label{sim-res}

Equation~(\ref{mass-trans2}) can be solved numerically using a forward 
finite-difference discretization of the time derivative. 
This leads to the following updating scheme: 

\begin{equation}
\label{mass-disc}
\tilde{m}(\tilde{\bf x},\tilde{t}_{k+1}) =   \tilde{m}(\tilde{\bf x},\tilde{t}_{k})
+ \delta t_{k} \,
\sum_{\tilde{\bf x'}} \, \left[ 
\tilde{W} (\tilde{\bf x'} \rightarrow \tilde{\bf x}, \tilde{t}_{k})
\, \tilde{m}(\tilde{\bf x'},\tilde{t}_{k})
 -   
\tilde{W} (\tilde{\bf x} \rightarrow \tilde{\bf x'}, \tilde{t}_{k}) \,
\tilde{m}(\tilde{\bf x},\tilde{t}_{k}) \right],
\end{equation}

\noindent
where the transition rate 
$\tilde{W} (\tilde{\bf x'} \rightarrow \tilde{\bf x}, \tilde{t}_{k})$
is given by 
$ \tilde{W} (\tilde{\bf x} \rightarrow \tilde{\bf x'}, \tilde{t}_{k})=
\exp \left[ - u \, \tilde{m}(\tilde{\bf x'},\tilde{t}_{k}) \right] $.
If $W_{\max,k} = \max_{{\bf x},{\bf x'}} \{ \tilde{W} (\tilde{\bf x} \rightarrow \tilde{\bf x'}, \tilde{t}_{k}), 
\tilde{W} (\tilde{\bf x'} \rightarrow \tilde{\bf x}, \tilde{t}_{k}) \}$ is 
the maximum transition rate at time $t_{k}$,  
the time increment $\delta t_{k}$ should satisfy the condition 
$\delta t_{k} \, \tilde{W}_{\max, k} <<1$.  
A very small time step would slow down the
evolution. Since the transition rates
change dynamically during the process, 
the time step is adaptively updated.

\begin{figure*}
\includegraphics[width=0.5\textwidth]{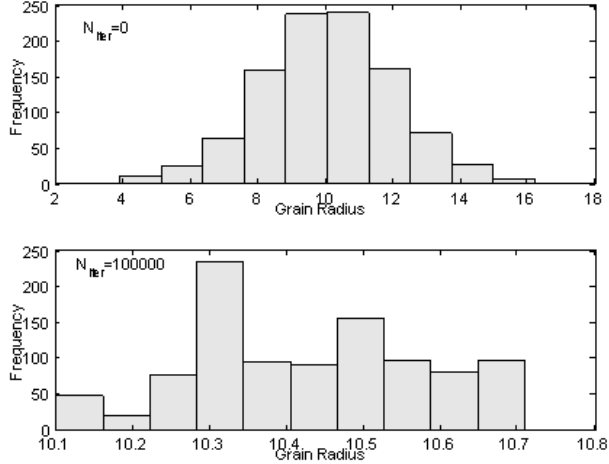}% Here is how to import EPS art
\caption{\label{Figure1} Grain size distribution at 
$t=0$ (top) and $t=100000$ (bottom) for $u=0.1$ and $\mu_{r,0}=0.2$.}
\end{figure*}

We solve Eq.~(\ref{mass-disc}) for a $1d$ chain of 
$N=1000$ grains with 
periodic boundary conditions. The initial grain distribution 
is assumed to be Gaussian with $\overline{r}_{0}=10$ in 
arbitrary units and coefficient of variation 
$\mu_{r,0}=0.2$.  
The reference mass $m_{0}$ is taken equal to 
the mean of the initial mass distribution. 
First, we consider the case of low activation energy
using a dimensionless activation parameter $u=0.1$. 
The initial grain size distribution and the 
evolved distribution after 100000 steps 
are shown in Fig.~(\ref{Figure1}).
In the evolved state the radius of all the grains is
approximately equal to the initial mean radius. 
In this case, the low activation energy leads to a diffusive
behaviour that asymptotically drives the system toward a
uniform steady state.

\begin{figure*}
\includegraphics[width=0.5\textwidth]{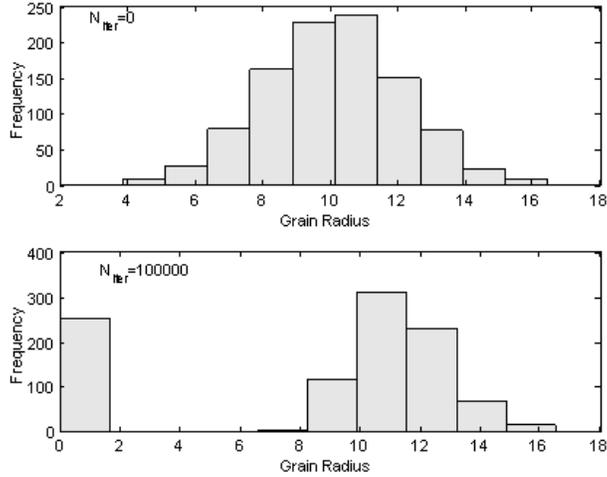}% Here is how to import EPS art
\caption{\label{Figure2} Grain size distribution 
for $u=15$ and $\mu_{r,0}=0.2$: initial 
(top) and after 100000 steps (bottom).}
\end{figure*}

\begin{figure*}
\includegraphics[width=0.5\textwidth]{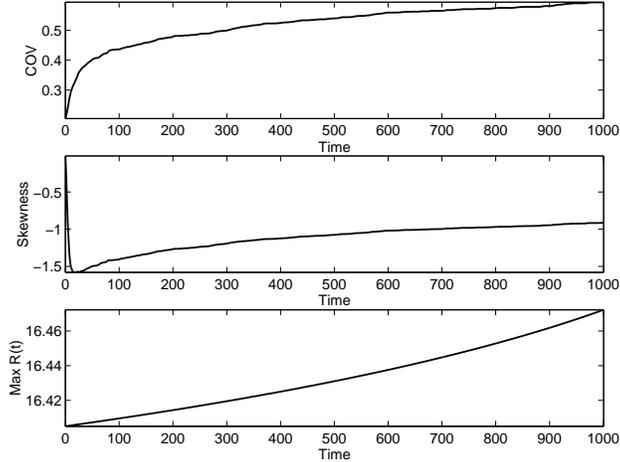}% Here is how to import EPS art
\caption{\label{Figure3} Coefficient of variation
 (top), skewness coefficient (middle), and 
 maximum radius (bottom) of the grain size distribution versus 
 simulation time for $u=15$ and $\mu_{r,0}=0.2$.}
\end{figure*}

\begin{figure*}
\includegraphics[width=0.5\textwidth]{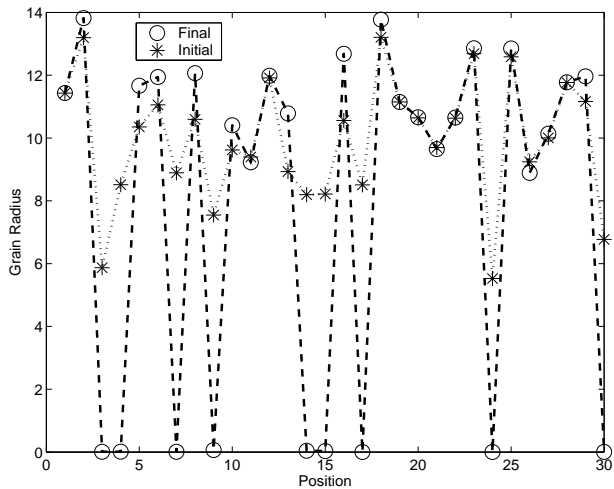}% Here is how to import EPS art
\caption{\label{Figure4} Initial (*) and final 
(o) grain sizes for the first 100 
sites; $u=15$ and $\mu_{r,0}=0.2$.}
\end{figure*}

Next, we consider the case of high activation energy. 
The grain size distribution
for $u=15$ is shown in Fig.~(\ref{Figure2}). 
Here, the larger grains grow at the expense
of the smaller ones. 
The grain radius distribution develops a bimodal 
structure that includes 
a fraction of very small grains.
The evolution of the grain radius 
coefficient of variation $\mu_{r}(t)$, skewness coefficient,
$s_{r}(t)$ and the 
maximum radius, $r_{max}(t)$ versus time are shown in 
Fig.~(\ref{Figure3}). The time is calculated based 
on $t_{k}=\sum_{i=0}^{k-1} \delta t_{i}$.
The coefficient of variation $\mu_{r}(t)$ 
increases in magnitude with time as a result of the evolving 
asymmetry of the distribution. 
The skewness coefficient $s_{r}(t)$ also develops a non-zero value as 
the distribution evolves away from the Gaussian. 
The $r_{max}(t)$ also increases 
albeit slowly. 
The plots shown in Fig.~(\ref{Figure3}) seem to indicate a
smooth approach to a time-independent steady state as the time increases.
Also note that if the evolution of the moments were plotted versus 
the number of steps (instead of the time $t_{k}$, the plots 
would exhibit discontinuities due to the fact that certain steps may 
correspond to significantly larger time increments than others.  
The spatial configuration of the grains is 
illustrated in Fig.~(\ref{Figure4}) by plotting the initial and final 
(after 100000 steps) radius of the first 30 grains: 
The smaller grains tend to shrink, while the larger 
grains grow at a considerably slower rate.
The asymmetry is due to mass conservation, i.e., 
the fact that an increase of the grain radius by $\delta r$
increases the mass by an amount 
that exceeds the respective reduction in mass due to a 
decrease $\delta r$ of the radius. A number of grains, the initial 
size of which is close to the mean radius, do not change appreciably
their size.

We note that the model presented here does not 
include grain coalescence. This can be addressed by solving 
Equation~(\ref{mass-disc}) 
iteratively, with the first iteration ending when the radius of 
one (or more) of the smaller grains drops below a certain 
threshold; at this point the smaller grains would coalesce with 
the largest nearest neighbours.  The number of grains would thus 
be reduced, and the resulting radius distribution would be used 
as an initial condition for the next iteration of Eq.~(\ref{mass-disc}). 
This mechanism leads to a coarse-graining of the 
grain radius distribution.  
The renormalization group approach may be a suitable 
framework for investigating the asymptotic grain size distribution 
under this coarse-graining procedure. 
 
\section{Conclusions}

We presented a discrete, nonlinear equation for grain growth 
in sintered grain aggregates, using transition rates that depend on the 
degree of amorphization of each grain. 
The model was solved numerically for
a  chain of grains with periodic boundary conditions.
The activation energy was shown to be a 
crucial factor for the asymptotic grain size distribution, 
since it leads to a transition from 
a bimodal steady state to a uniform one 
(diffusion regime). Various aspects of the 
model require further 
study, including the existence of a well-defined 
threshold between the two regimes, 
 the impact of governing 
parameters on the grain size evolution, the roles
of the initial grain size distribution and the 
initial grain configuration.  
Finally, grain coalescence needs to 
be incorporated in the model (in the spirit 
discussed in the previous section) in order to observe 
realistic grain growth.

% \appendix

% \section{}
% \label{}

\section*{Acknowledgments}
Research supported by the
EC grant `Super High Energy Milling in the Production of Hard Alloys, 
Ceramic and Composite Materials' (Activation FP6).

\end{document}